\documentstyle[12pt]{article}

\begin{document} 
\title{PATH INTEGRAL AND THE INDUCTION LAW}
\author{{\Large F.A. Barone $^{\star}$, C. Farina$^{\dagger}$} \\\\
Instituto de F\'{\i}sica - UFRJ - CP 68528
\\
Rio de Janeiro, RJ, Brasil - 21945-970.}
\maketitle
\begin{abstract}
{We show how the induction law is correctly used in the path integral computation of the 
 free particle propagator. The way this primary path integral example is treated in most textbooks is  a little bit missleading.}
\end{abstract}

\bigskip
\vfill
\noindent $^{\star }$ {e-mail: fabricio@if.ufrj.br}

\noindent $^\dagger$ {e-mail: farina@if.ufrj.br}

\pagebreak

The path integral quantization method was developed in detail by Feynman \cite{Feynman48} in 1948. Feynman developed some earlier ideas introduced by Dirac \cite{Dirac33}. Since then, path integral methods have provided a good alternative procedure to quantization and many books have been written on this subject, not only in quantum 
mechanics \cite{FeynmanHibbs,Schulman,Khandekar,Kleinert,Inomata}, but also in quantum field theory \cite{Fried,Rivers,Das} (many modern textbooks in quantum field theory devote a few chapters to functional methods). We can surely say that in the last decades, Feynman's method  has been recognized as a very convenient and economic mathematical tool for treating problems in a great variety of areas in physics, from ordinary quantum mechanics and statistical quantum mechanics to quantum field theory and condensed matter field. It is a common feature of almost all texts which introduce the Feynman quantization  prescription  to use the unidimensional free particle propagator as a first example. In many cases, this simple example is the only one that is explicitly evaluated. The reason for that is simple: after the free particle propagator has been presented, it is usual to introduce the semiclassical method, which is exact for quadratic lagrangians, so that examples like the oscillator propagator or the propagator for a charged particle in a uniform magnetic field can be obtained without the explicit calculation of the Feynman path integral (for the oscillator propagator, the reader may find both calculations, that is, the explicit one and the semiclassical one in Ref.\cite{Holstein}; see also references therein). Curious as it may seem, the free particle propagator is not treated as it should,  regarding the correct use of the mathematical induction law. It is the purpose of this note to show how the induction law shold be applied to the free  particle propagator in the context of path integrals. In what follows, we first make some comments about the usual way of obtaining this propagator and then we show how one should proceed if the use of induction law is taken seriously. 

The Feynman prescription for the quantum mechanical transition amplitude $K(x_{N},x_{0};\tau)$ of a particle which was localized at $x_{0}$ at time $t=0$, to be at the position $x_{N}$ at time $t=\tau$ (called Feynman propagator) is given by the path integral \cite{FeynmanHibbs}:
\begin {eqnarray}\label{1}
K(x_{N},x_{0};\tau)&=&\lim_{N\rightarrow\infty\atop{N\varepsilon=\tau}}{\sqrt{m\over 2\pi i\hbar\varepsilon}}\int\prod_{j=1}^{N-1}\biggl({\sqrt{m\over 2\pi i\hbar\varepsilon}dx_{j}}\biggr)\times\nonumber\\\nonumber\\
&\times&
\exp\Biggl\{{i\over\hbar}\sum_{k=1}^{N}\Biggl[{m(x_{k}-x_{k-1})^{2}\over 2\varepsilon}-\varepsilon V\Biggl({x_{k}+x_{k-1}\over 2}\Biggr)\Biggr]\Biggr\}\; ,
\end {eqnarray}
where $V(x)$ is the potential energy of the particle. Setting $V(x)=0$ in the above equation, we get the free particle Feynman propagator:

\begin {eqnarray}
\label {propagadorintermediario1}
K(x_{N},x_{0};\tau)&=&\lim_{N\rightarrow\infty\atop{\varepsilon\rightarrow 0}}
{\sqrt{m\over 2\pi i\hbar\varepsilon}}
\int\biggl({\sqrt{m\over 2\pi i\hbar\varepsilon}dx_{N-1}}\biggr)...\int\biggl({\sqrt{m\over 2\pi i\hbar\varepsilon}dx_{2}}\biggr)\times\nonumber\\\nonumber\\
&\times&{\sqrt{m\over 2\pi i\hbar\varepsilon}}\int\biggl({\sqrt{m\over 2\pi i\hbar\varepsilon}dx_{1}}\biggr)\exp\Biggl[{im\over 2\hbar\varepsilon}\sum_{k=1}^{N}(x_{k}-x_{k-1})^{2}\Biggr]
\end {eqnarray}
For convenience, let us define 
\begin{equation}\label{Ij}
I_j(x_0,x_{j+1}):= \left(\sqrt{{m\over 2\pi i\hbar\varepsilon}}\;\right)^{j+1}
\int_{-\infty}^\infty\, dx_j\; ...\int_{-\infty}^\infty\, dx_1\;
\exp\left[{im\over 2\hbar\varepsilon}\sum_{k=1}^{j+1}\left( x_k-x_{k-1}\right)^2
\right]\; ,
\end{equation}
where $j=1,2,...,N-1$, so that $I_1$ corresponds to the result of the first integration (with two normalization factors taken into account), $I_2$ corresponds to the result of the first two integrations (with three normalization factors taken into account), etc.. As a consequence of the previous definition, we can write:
\begin{equation}\label{recorrencia}
I_{j+1}(x_0,x_{j+2})=\int\, dx_{j+1}\;{\sqrt{m\over 2\pi i\hbar\varepsilon}}\; 
\exp\Biggl[{im\over 2\hbar\varepsilon}(x_{j+2}-x_{j+1})^{2}\Biggr]I_{j}(x_0,x_{j+1})
\end{equation}
and it is also clear that:
%
\begin {equation}
\label {propagadorfuncaodeIj}
K(x_{N},x_{0};\tau)=\lim_{N\rightarrow\infty}I_{N-1}(x_0,x_N)\; .
\end {equation}

What is usually done in the literature is the following: one firstly obtains the expression for $I_1$, which can be done by completing the square in the argument of the exponential of the integrand, that is,
\begin {eqnarray}
\label {resultadoI1}
I_{1}(x_0,x_2)&=&{m\over 2\pi i\hbar\varepsilon}\exp{\biggl[{im\over 2\hbar}{(x_{2}-x_{0})^{2}\over 2\varepsilon}\biggr]}\int_{-\infty}^{\infty}dx_{1}\exp\biggl\{{im\over\hbar\varepsilon}\biggl[x_{1}-\biggl({x_{0}+x_{2}\over 2}\biggr)\biggr]^{2}\biggr\}\nonumber\\\nonumber\\
&=&{\sqrt{m\over 2\pi i\hbar(2\varepsilon)}}\;
\exp\biggl[{im\over 2\hbar}{(x_{2}-x_{0})^{2}\over 2\varepsilon}\biggr]\; ,
\end {eqnarray}
where we have used the Fresnel integral \cite {Arfken}. Next, using Eq.(\ref{recorrencia}) and the above result for $I_1(x_0,x_2)$, one proceeds and obtains the expression for $I_2$:
\begin {eqnarray}
\label {resultadoI2}
I_{2}(x_0,x_3)&=&\sqrt{m\over 2\pi i\hbar\varepsilon}{\sqrt{m\over 2\pi i\hbar(2\varepsilon)}}\int_{-\infty}^{\infty}dx_{2}\exp\Biggl\{{im\over 2\hbar\varepsilon}\Biggr[(x_{3}-x_{2})^{2}+{1\over 2}(x_{2}-x_{0})^{2}\Biggr]\Biggr\}\nonumber\\\nonumber\\
&=&{\sqrt{m\over 2\pi i\hbar(3\varepsilon)}}\exp\biggl[{im\over 2\hbar}{(x_{3}-x_{0})^{2}\over 3\varepsilon}\biggr]\; .
\end {eqnarray}
The last two formulas strongly suggest that after $j$ integrals have been evaluated, the result of $I_j$ is given by:
%
\begin {equation}
\label {Ijsuposto}
I_{j}(x_0,x_{j+1})={\sqrt{m\over 2\pi i\hbar(j+1)\,\varepsilon}}\;
\exp\biggl[{im\over 2\hbar}{(x_{j+1}-x_{0})^{2}\over (j+1)\varepsilon}\biggr]\; .
\end {equation}
It is common to accept that Eqs.(\ref{resultadoI1}) and (\ref{resultadoI2}) are sufficient to demonstrate Eq.(\ref{Ijsuposto}), so that the final expressions for the desired propagator is given by:
\begin{eqnarray}\label{propagfinal1}
K(x_N,x_0;\tau)&=&\lim_{N\rightarrow\infty\atop{\varepsilon\rightarrow 0}}\; I_{N-1}(x_0,x_N)\nonumber\\
\nonumber\\
&=& \lim_{N\rightarrow\infty\atop{\varepsilon\rightarrow 0}}\;
\left\{\sqrt{{m\over 2\pi i\hbar(N\varepsilon)}}\;
\exp\left[ {im\over 2\hbar}{(x_N-x_0)^2\over (N\varepsilon)}\right]\right\}\nonumber\\
\nonumber\\
&=& \sqrt{{m\over 2\pi i\hbar\tau}}\;
\exp\left[ {im\over 2\hbar}{(x_N-x_0)^2\over \tau}\right]\; ,
\end{eqnarray}
which is, in fact, the correct answer.

However, a \lq\lq strongly suggested result{\rq\rq} is not enough to be considered as a mathematical demonstration of a result. A rigorous demonstration of Eq.(\ref{Ijsuposto}) for $j=1,2,...,N-1$ requires the correct use of the mathematical induction law, which we pass now to discuss.

To apply correctly the induction law to the problem at hand means the following: we first demonstrate the validity of Eq.(\ref{Ijsuposto}) for $j=1$ and then we demonstrate that if this equation is true for an arbitrary $j$, it will also be true for $j+1$. The first step is already done, see Eq.(\ref{resultadoI1}). To complete the demonstration, let us assume that Eq.(\ref{Ijsuposto}) is valid for an arbitrary $j$. Therefore, using Eq.(\ref{recorrencia}) the expression for $I_{j+1}$ is given by:
\begin {eqnarray}
I_{j+1}(x_0,x_{j+2})&=&{\sqrt{m\over 2\pi i\hbar\varepsilon}}{\sqrt{m\over 2\pi i\hbar(j+1)\varepsilon}}
\times\nonumber\\
\nonumber\\
&\times& \int_{-\infty}^{\infty}dx_{j+1}\exp\biggl\{{im\over 2\hbar(j+1)\varepsilon}[(j+1)(x_{j+2}-x_{j+1})^{2}+(x_{j+1}-x_{0})^{2}]\biggr\}
\nonumber\\
\end {eqnarray}
Noting that:
\begin {eqnarray}
(j+1)(x_{j+2}-x_{j+1})^{2}+(x_{j+1}-x_{0})^{2}&=&\biggl({j+1\over j+2}\biggr)(x_{j+2}-x_{0})^{2}\nonumber\\
\nonumber\\
&+&(j+2)\biggl\{x_{j+1}-{1\over j+2}[(j+1)x_{j+2}+x_{0}]\biggr\}^{2}\nonumber\\
\end {eqnarray}
we have:
\begin {eqnarray}
I_{j+1}(x_0,x_{j+2})&=&{m\over 2\pi i\hbar\varepsilon}{1\over\sqrt{j+1}}\exp\biggl\{{im\over 2\hbar(j+2)\varepsilon}(x_{j+2}-x_{0})^{2}\biggr\}\times\cr\cr\cr
&\times&\int_{-\infty}^{\infty}dx_{j+1}\exp\biggl\{{im(j+2)\over 2\hbar\varepsilon(j+1)}\biggl[x_{j+1}-{(j+1)x_{j+2}+x_{0}\over j+2}\biggr]^{2}\biggr\}\cr\cr\cr
&=&{\sqrt{m\over 2\pi i\hbar[(j+1)+1]\varepsilon}}\;
\exp\biggl\{{im\over 2\hbar}{(x_{(j+1)+1}-x_{0})^{2}\over [(j+1)+1]\varepsilon}\biggr\}\; ,\\
\nonumber
\end {eqnarray}
which is precisely Eq.(\ref{Ijsuposto}) if we replace in this equation $j$ by $j+1$. Hence, we have succeeded in demonstrating that the validity of this equation for an arbitrary $j$ implies indeed its validity for $j+1$ and as a consequence, Eq.(\ref{propagfinal1}) is now rigorously justified. Though this is  the simplest quantum propagator, it is in general the first example presented by most texts in path integral quantization and we think that if it is done with a reasonable mathematical rigor it is a good beginning for those who intend to step into the path integral world.

\bigskip
\noindent
{\bf Acknowledgments:} the authors are indebted with M.V. Cougo-Pinto and A.C. Tort for reading the manuscript. C.F. and F.B. would like to thank CNPq and CAPES, respectively, for partial financial support.

\vfill\eject
\noindent 

\end {document}